\documentclass[%
 reprint,
superscriptaddress,
 amsmath,amssymb,
 aps,
]{revtex4-2}

\usepackage{graphicx}
\usepackage{dcolumn}
\usepackage{bm}
\usepackage{amsmath}
\usepackage{comment}
\usepackage{hyperref}
\usepackage{color}

\begin{document}

\preprint{APS/123-QED}

\title{Transonic Solutions for Recombination-Driven Stellar Winds}

\author{Ilan Strusberg}
\email{Ilan.Strusberg@mail.huji.ac.il}
\affiliation{Racah Institute of Physics, The Hebrew University of Jerusalem, 9190401, Israel}

\author{Re'em Sari}
\affiliation{Racah Institute of Physics, The Hebrew University of Jerusalem, 9190401, Israel}

\author{Jim Fuller}
\affiliation{TAPIR, Mailcode 350-17, California Institute of Technology, Pasadena, CA 91125, USA}
\date{\today}

\begin{abstract}

We present an analytical stationary isentropic solution of the spherically symmetric Euler equations in the gravitational field of a star using an equation of state of ionizable monatomic gas. The solution consists of
a fully ionized hydrostatic inner region, followed by a thick hydrostatic recombination region where the density decreases by orders of magnitude, the radius increases by about an order of magnitude and the temperature decreases by a factor of two. Within this recombination region, a large portion of the recombination energy is used for lifting the gas subsonically. This region ends at a critical point, located roughly at $10-100~\rm{AU}$, where the gas is mostly recombined, beyond which it flows supersonically as a wind. We find the position and the quantities of the gas at the critical point and derive the mass-loss rate of the solution. 

We apply our solution to evolved stars, with a compact core surrounded by a high entropy envelope.
We derive the mass-loss rate as a function of time. As the star is losing mass, it goes through a sequence of our solutions, in a runaway manner which ends once most of the high entropy envelope is lost. The recombination-driven winds are initiated once the stars expands to a radius of about $1~\rm{AU}~M/M_\odot$ and are terminated on a timescale of $10^1-10^4~\rm{years}$. We discuss the implications for common envelope evolution.
\end{abstract}

\maketitle

\section{Introduction}
\label{sec: Introduction}
In his famous work, Parker analytically modeled the solar wind emitted by the Sun \cite{parker1958dynamics, Parker1965SolarWind}. Parker derived a transonic solution to the spherically symmetric Euler equations within the gravitational field of a star, and proved that there exists a unique solution where the gas transitions smoothly from subsonic to supersonic speeds. The point where the flow shifts from being near hydrostatic to an outflowing wind with a nearly constant velocity is referred to as the \emph{Critical Point}.
In his well-known solution, Parker assumed the outflowing gas is isothermal, with a characteristic temperature of $T \simeq 1 \cdot 10^6~\rm{K}$. Several studies have revisited the solar wind solution under slightly different assumptions \cite{Parker1972_PresentDevelopmentsSolarWind,holzer1979solar, keppens1999numericalsimulationsstellarwinds, shivamoggi2021parker, shivamoggi2024polytropicgaseffectsparkers}, most notably by considering polytropic outflows. These works reproduce Parker’s key result, that given the boundary conditions at the stellar surface, there exists a unique stationary transonic solution.

In this paper, we investigate the hydrodynamics of ionized gas which is either in the ionized state or the ground state, with a particular focus on how the recombination of the atoms and free electrons influences the gas flow. We assume the gas obeys the following equation of state (EOS):
\begin{equation}
    P=\frac{k_BT\rho}{\mu}\left(1+\eta\right),
    \label{eq: EOS of Ionized Gas}
\end{equation}
where $k_B$ is the Boltzmann constant and $\mu$ is the mean mass per baryon. $P$, $T$ and $\rho$ are the pressure, temperature and density of the gas, respectively. The ionization fraction, $\eta$, is the ratio of the number of atoms in the ionized state to the total number of atoms.

Although a transonic solution does not exist for a polytropic flow of an ideal gas with an adiabatic index of \( \gamma = 5/3 \) \cite{holzer1979solar}, such a solution arises for the EOS given in Eq. (\ref{eq: EOS of Ionized Gas}). We show that this solution can be characterized by two distinct regions: subsonic adiabatic flow that connects smoothly to the underlying hydrostatic structure of the star, overlaid by a wind solution with a nearly constant hypersonic velocity. Between the two regions, the gas accelerates from subsonic to hypersonic speeds, passing through the transonic critical point. During this transition, the ionization fraction decreases with temperature, evolving from a nearly fully ionized state at the base of the wind, to a completely recombined outflow. Although descriptions of ionized-gas stellar winds already exist \cite{newman1968recombination, Waldron1984, Williams_1996}, these studies address different physical scenarios, and they neglect the gravitational field of the source, which plays a crucial role in the current work. 
There is a similarity between our solutions to the ones presented by \cite{Quataert_2016}, we discuss this in \S\ref{sec: The Critical Point}.

The ionization energy has been proposed as an important energy source for the envelope ejection during \emph{Common Envelope Interaction} (CEI) \cite{Reichardt_2020, Ivanova_2015, Nandez_2016, Nandez_2015, Ivanova2016Common}. In these events, the companion in a binary system becomes engulfed by the envelope of the donor, forming a common envelope that may ultimately result in the merger of the two stars or in the complete ejection of the envelope. Such events have been extensively studied in the literature, analytically and numerically \cite{Paczynski_1976, Webbink1984, Livio1988, DeMarco2011,  CommonEnvelopeEvolutionBook, Ivanova_2013, Ivanova_2016, Ricker_2007, Lau_2022_A, Lau_2022_B, refId0, roepke2022simulationscommonenvelopeevolutionbinary, Nelemans2005, Di_Stefano_2023}. Numerical simulations that include the ionization energy show the existence of recombination outflows \cite{Ivanova_2015, Nandez_2015}, which are analyzed analytically in the present work. However, the efficiency with which recombination energy contributes to the envelope ejection is still debated, mainly due to radiation leakage from the envelope and additional physical effects \cite{Sabach_2017, chamandy2024negativefeedbackambientenvironment, Grichener_2018, Soker_2018, Ivanova_2018}. In this paper, we neglect such effects, and their impact on our results should be carefully examined in future studies. Several studies that include radiation effects, link between CEI and luminous red novae (LRNe), and attempt to model these transients within this framework \cite{chen2024bridginggapluminousred, Kirilov_2025}.

In \S\ref{sec: Adiabatic Expansion}, we describe the adiabatic expansion of ionized gas obeying the EOS given by Eq. (\ref{eq: EOS of Ionized Gas}), and discuss its behavior in the limit of high entropy per baryon. In \S\ref{sec: The Transonic Solution}, we describe the transonic wind solution for the flow of the gas during recombination: In \S \ref{sec: Hydrostatic Equilibrium} we find the ionization fraction profile in the high entropy limit- and consequently, all other physical quantities of the gas below the critical point (in the hydrostatic equilibrium region), and in \S\ref{sec: The Critical Point} we find the position of the critical point and the corresponding mass-loss rate. In \S \ref{sec: Evolution of Ionized Envelope}, we discuss the evolution of the polytropic envelope of the star that is the origin of the recombination-driven winds, and show they lead to a runaway process. In \S \ref{sec: Recombination-Driven Winds in CEI} we discuss how recombination-driven winds would cause full envelope ejection during CEI. In \S\ref{sec: Conclusions and Discussion} we summarize our results and discuss their possible implications.

\section{Adiabatic Expansion of Ionized Gas}\label{sec: Adiabatic Expansion}
In this section, we will examine the adiabatic expansion of a single-level ionized gas, initially in a nearly fully ionized state ($\eta\approx1$). 

We focus on hydrogen gas, which is either in the ground state or the ionized state, with degeneracies of $g_0=1$ and $g_1=2$, respectively.  
In this case, the mean mass per baryon is $\mu=1.67\cdot10^{-24}g$ and the ionization energy is $U=13.6~eV$, which is equivalent to $1.58\cdot10^5K$. We determine the ionization fraction of the gas $\eta$, defined as the ratio of the number density of atoms in the ionized state to the total number density, as a function of the temperature $T$ and the density $\rho$, using the \emph{Saha Equation} \cite{Saha1920, Saha1921} for hydrogen,
\begin{equation}
    \frac{\eta^2}{1-\eta}=\frac{\mu}{\lambda_{\rm{e}}^3\rho}\exp{\left(-\frac{U}{k_BT}\right)},
    \label{eq: Saha Equation with one ionization level}
\end{equation}
where
\begin{equation}
    \lambda_{\rm{e}}\equiv\frac{h}{\sqrt{2\pi m_ek_BT}}
    \label{eq: The thermal de-Broglie wavelength}
\end{equation}
is de-Broglie wavelength of electrons in thermal equilibrium with a temperature $T$, with $m_e$ being the electron mass.
The entropy per baryon $S$, as a function of the temperature $T$ and the ionization fraction $\eta$, presented in \cite{corli2017ionizedgasdynamics}, is given by
\begin{equation}
    S=\left(\frac{5}{2}+\frac{U}{k_BT}\right)\left(1+\eta\right)+2\ln\left(\frac{\eta}{1-\eta}\right)+S_0,
    \label{eq: Entropy of Ionized Gas}
\end{equation}
where $S_0\equiv1.5\ln{\left(\mu/m_e\right)}+\ln{2}=12.0$ is a constant required for Eq. (\ref{eq: Entropy of Ionized Gas}) to converge to the Sackur-Tetrode formula in the limit of non-ionized gas, i.e. $\eta\rightarrow0$ (throughout the paper, the entropy per baryon of the gas $S$ is given in units of $k_B$). 
From here on, we will use $\Delta S\equiv S-S_0$ for convenience. 
Substituting Eq. (\ref{eq: Entropy of Ionized Gas}) into Eq. (\ref{eq: Saha Equation with one ionization level}), we determine the density profile during adiabatic expansion, 
\begin{equation}
    \rho=\frac{\mu}{\lambda_{\rm{e}}^3}F(\eta)\exp{\left(\frac{5}{2}-\frac{\Delta S}{1+\eta}\right)},
    \label{eq: Density Profile During Adaiabatic Expansion}
\end{equation}
where
\begin{equation}
    F(\eta)\equiv \eta^{-\frac{2\eta}{1+\eta}}\left(1-\eta\right)^{-\frac{1-\eta}{1+\eta}}
\end{equation}
is a dimensionless function, that approaches $F(\eta)\rightarrow1$ in the limits $\eta\rightarrow0$ and $\eta\rightarrow1$.
Due to this, in the fully ionized and non-ionized limits, the density profile during adiabatic expansion reduces to the well-known polytropic relations, i.e. $\rho\propto T^{3/2}$.
Using the results above, we derive the adiabatic index, defined as $\gamma\equiv\left(\frac{\partial \ln{P}}{\partial\ln{\rho}}\right)_S$, as a function of the temperature and the ionization fraction $\eta$,
\begin{equation}
    \gamma=\frac{\left(5+6.25\tilde{T}+\tilde{T}^{-1}\right)\eta\left(1-\eta\right)+5\tilde{T}}{\left(3+3.75\tilde{T}+\tilde{T}^{-1}\right)\eta\left(1-\eta\right)+3\tilde{T}},
    \label{eq: The adiabatic index of inozed gas}
\end{equation}
where $\tilde{T}\equiv k_BT/U$. For completely ionized and non-ionized gas, it approaches the known value for monatomic ideal gas, $\gamma\approx5/3$. The sound speed, calculated from the adiabatic index, is given by $c_s^2\equiv\gamma P/\rho$.
Assuming $S\gg1$, Eq. (\ref{eq: Entropy of Ionized Gas}) predicts that the gas, initially nearly fully ionized, only begins its recombination when its temperature becomes of order of $T_{\rm{ion}}=\Delta S^{-1} U/k_B$, which we refer to as \emph{The Ionization Temperature}. Below the ionization temperature, the gas is almost completely non-ionized, i.e. $\eta\ll1$. In the limit $S\rightarrow\infty$, the gas remains nearly fully ionized until the temperature drops below twice the ionization temperature. Below this threshold, the ionization fraction $\eta$ decreases roughly linearly with temperature, $\eta(T)\simeq T/T_{\rm{ion}}-1$, until it reaches the ionization temperature.
Notice that in this limit, although the temperature only changes by a factor of two, Eq. (\ref{eq: Density Profile During Adaiabatic Expansion}) implies that the density and pressure decrease by a factor of $\exp{\left(\Delta S/2\right)}\gg1$ (many orders of magnitude). 
Consequently, in this limit, the adiabatic index of a partially ionized gas (given by Eq. \ref{eq: The adiabatic index of inozed gas}) approaches $\gamma\rightarrow1$.

\begin{figure}[hbtp]
    \centering
    \includegraphics[width=\linewidth]{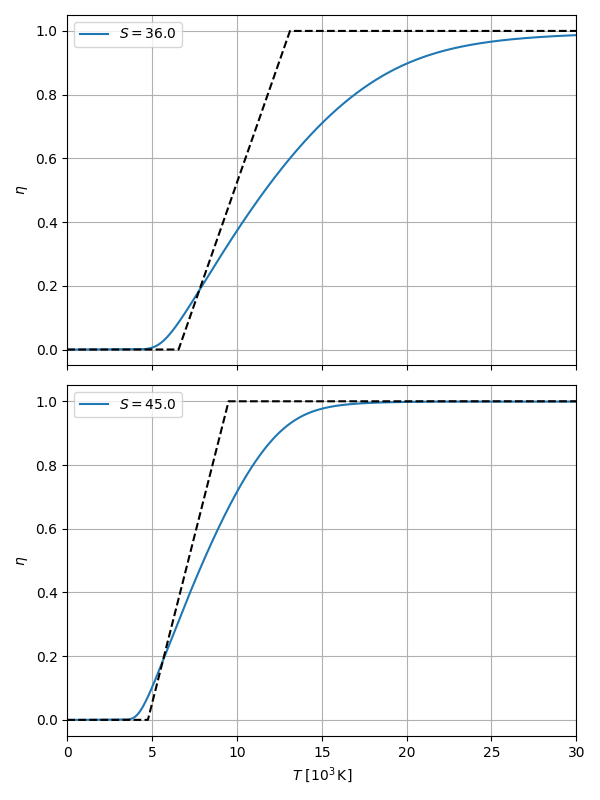}
    \caption{The ionization fraction of hydrogen gas undergoing adiabatic expansion, plotted as a function of temperature (in units of $10^3K$). The top and bottom panels correspond to entropies per baryon of $S=36.0$ and $S=45.0$, respectively (corresponding to $\Delta S=24.0$ and $\Delta S = 33.0$). These entropies correspond to polytropic envelope masses of $1~M_\odot$ and $10~M_\odot$ (see Eq. \ref{eq: Stellar Mass Function} for more details).
    The solid blue lines show the ionization fraction given by the Saha equation (Eq. \ref{eq: Entropy of Ionized Gas}), while the black dashed lines show the ionization fraction given by the approximation of the high entropy limit: In the regime $T > 2T_{\rm{ion}}$, the gas is completely ionized ($\eta=1$), in the regime $T < T_{\rm{ion}}$ the gas is non-ionized ($\eta=0$), and between them the ionization fraction depends linearly on the temperature, i.e. $\eta=T/T_{\rm{ion}}-1$. The maximal mismatch between the Saha equation and the high entropy approximation, that scales as $\Delta S^{-1}$, reaches $40\%$ for $S=36.0$ and $33\%$ for $S=45.0$.}
    \label{fig: Adiabatic Expansion}
\end{figure}

Fig. (\ref{fig: Adiabatic Expansion}) presents the ionization fraction as a function of temperature, as given by Eq. (\ref{eq: Entropy of Ionized Gas}), for entropies per baryon of $S=36.0$ and $S=45.0$ (corresponding to $\Delta S=24.0$ and $\Delta S = 33.0$, respectively). According to Eq. (\ref{eq: Stellar Mass Function}), these entropies correspond to polytropic envelope masses of $1~M_\odot$ and $10~M_\odot$, respectively. Fig. (\ref{fig: Adiabatic Expansion}) also shows the ionization fraction given by the approximation of the high entropy limit, calculated using the appropriate ionization temperatures.

As illustrated by Fig. (\ref{fig: Adiabatic Expansion}), the mismatch between the ionization fraction given by the Saha equation (Eq. \ref{eq: Entropy of Ionized Gas}) and the one given by the limit of high entropy is significant for such entropies, reaching $~40\%$ at its peak for an entropy per baryon of $S=36.0$. While  this approximation captures the correct order of magnitude of all physical quantities during the adiabatic expansion, it introduces a non-negligible error in the calculation of each of them. We emphasize that, in the present work, we employ the high entropy limit solely to demonstrate the behavior of recombination-driven winds to first order in $\Delta S^{-1}$, and that more accurate results can be obtained numerically (see appendix \ref{app: The Accuracy of The High Entropy Limit} for more details). 

\section{The Transonic Solution}
\label{sec: The Transonic Solution}
We aim at finding a transonic stationary solution describing the flow of gas that obeys the EOS of ionized gas (Eq. \ref{eq: EOS of Ionized Gas}) in the gravitational field of a star of mass $M$. The solution describes gas that accelerates from subsonic to hypersonic speeds, and it transitions smoothly between the hydrostatic and wind regimes. This continuity implies the existence of a critical point, where the local velocity of the gas, $v_c$, is equal to the local sound speed, $c_{s,c}$.
In appendix \ref{app: Uniqness of The Transonic Solution}, we demonstrate that the stationary solution is determined by requiring it to be differentiable at the critical point. We use this to determine the position of the critical point and the properties of the gas when it reaches it. We then use these results to calculate the corresponding mass-loss rate of the star due to recombination-driven winds, which uniquely determines the solution everywhere.
Since we assume the flow of the gas is adiabatic, the entropy per baryon of the gas, $S$, is constant along the flow. For simplicity, we assume the star has a polytropic envelope, implying that the specific entropy is independent of the radius. This assumption is reasonable if the star has an outer convective zone, which contains most of the envelope's mass, as is the case for massive red giants.
Deep inside the star, the gas is nearly fully ionized. We work under the assumption that $S\gg1$, and calculate all physical quantities to an order of $\Delta S^{-1}$.
As discussed in \S \ref{sec: Adiabatic Expansion}, this assumption implies that when the gas starts its recombination, its density drops drastically.
Thus, we do not consider the effect of the outflowing partially ionized gas
on the gravitational field outside the point where its temperature drops below twice the ionization temperature (see Eq. \ref{eq: Recombination Radius} and the discussion below), since its mass (enclosed by the radius of the critical point) is expected to be negligible.

The flow of the gas is governed by the stationary spherically symmetric Euler equations in a gravitational field, given by:
\begin{align}
    \label{eq: Mass conservation Euler Equation}
    0&=\frac{1}{r^2}\partial_r\left(\rho vr^2\right),\\
    \label{eq: Momentum Conservation Euler Equation}
    0&=v\partial_rv+\frac{1}{\rho}\partial_rP+\frac{GM}{r^2},\\
    \label{eq: Energy Conservation Euler Equation}
    0&=\frac{1}{r^2}\partial_r\left(\rho vr^2\left(\frac{v^2}{2}+e+\frac{P}{\rho}-\frac{GM}{r}\right)\right),
\end{align}
where $\rho$, $v$ and $P$ are the density, velocity, and pressure of the gas, respectively. All time derivatives were dropped since we seek the stationary solution. The specific energy of a single-level partially ionized gas $e$ is given by
\begin{equation}
    e=\frac{3}{2}\frac{k_BT}{\mu}\left(1+\eta\right)+\frac{U}{\mu}\eta.
    \label{eq: The Specific Energy of Ionized Gas}
\end{equation} 
By integrating Eqs. (\ref{eq: Mass conservation Euler Equation}) and (\ref{eq: Energy Conservation Euler Equation}), we find two quantities 
independent of $r$,
\begin{align}
    \label{eq: mass-loss Rate}
    \dot M&=4\pi r^2\rho v,\\
    \varepsilon&=\frac{v^2}{2}+ \frac{5}{2}\frac{k_BT}{\mu}\left(1+\eta\right)+\frac{U}{\mu}\eta-\frac{GM}{r}.
    \label{eq: Bernoulli Parameter of Transonic Solution}
\end{align}
The first quantity, $\dot M$, is the mass-loss rate of the star due to recombination-driven winds, and the second one, $\varepsilon$, is the \emph{Bernoulli Parameter} of the flow. These two properties, together with the entropy per baryon $S$, determine the solution at any radius.
Similarly to the entropy per baryon $S$, the Bernoulli parameter of the solution $\varepsilon$ is also determined from the properties of the polytropic envelope of the  star. While the sum of the specific enthalpy and the gravitational potential
is generally not constant inside the star, it is conserved within regions of constant specific entropy, as in polytropic envelopes (in stationary flows, the gradient of the Bernoulli constant is the gradient of the entropy times the temperature). Thus, the Bernoulli parameter of the transonic solution is equal to the sum of the specific enthalpy and the gravitational potential in the polytropic envelope. In appendix \ref{app: Uniqness of The Transonic Solution} we show how to derive the mass-loss rate directly from the entropy and Bernoulli parameters. Thus, the transonic solution depends solely on the intrinsic properties of the gas deep inside the hydrostatic region.

\subsection{Hydrostatic Equilibrium}
\label{sec: Hydrostatic Equilibrium}
As explained earlier, deep inside the star, the gas is assumed to be nearly fully ionized, i.e. $\eta\approx1$. We follow a gas element, that flows outward from the fully ionized region to the critical point. As the gas element flows outward, $\eta$ decreases, and according to Eq. (\ref{eq: Density Profile During Adaiabatic Expansion}) its density drops drastically. 
Consequently, the velocity of the gas element increases significantly (Eq. \ref{eq: mass-loss Rate}). The temperature of the gas element remains nearly constant (of order the ionization temperature, $T_{\rm{ion}}=\Delta S^{-1}U/k_B$), until the gas fully recombines, which only occurs after the gas element crosses the critical point. 
As discussed in \S \ref{sec: Adiabatic Expansion}, in the region where the gas is only partially ionized, the specific thermal energy is of order $k_BT_{\rm{ion}}/\mu\simeq\Delta S^{-1}U/\mu$, which is much lower than the specific ionization energy $U/\mu$. Note that this is not true deep inside the star, where the gas is nearly fully ionized and the temperature reaches above $\sim10^5 \,$K. At radii below the critical point, the kinetic energy is also negligible compared to the ionization and potential energies. Thus, we extract the dependence of the ionization fraction on the radius in the hydrostatic equilibrium regime from Eq. (\ref{eq: Bernoulli Parameter of Transonic Solution}), by neglecting all energies except the potential and ionization energies:
\begin{equation}
    \eta(r)\approx \frac{\varepsilon+GM/r}{U/\mu}.
    \label{eq: Ionization Fraction Before Critical Point}
\end{equation}
Consequently, we deduce the dependence of the temperature and density on the radius by substituting the result above into Eqs. (\ref{eq: Entropy of Ionized Gas}) and (\ref{eq: Density Profile During Adaiabatic Expansion}), respectively. Eq. (\ref{eq: Ionization Fraction Before Critical Point}) implies the gas is partially ionized only for radii above
\begin{equation}
    r_{\rm{rec}} = \frac{GM}{U/\mu-\varepsilon}= 0.68~\frac{M}{M_\odot}\left(1-\frac{\mu\varepsilon}{U}\right)^{-1}\rm{AU}.
    \label{eq: Recombination Radius}
\end{equation}
This radius, referred to as \emph{The Recombination Radius}, is defined as the point where the temperature is approximately twice the ionization temperature $T_{\rm{ion}}$. In the high entropy limit, this radius marks the point where the gas begins its recombination. Initially, most of the stellar mass is found below this radius. The treatment of this part of the solution is dealt with in \S \ref{sec: Evolution of Ionized Envelope}. 

\subsection{The Critical Point}
\label{sec: The Critical Point}
We use the results of appendix \ref{app: Uniqness of The Transonic Solution}, which is a generalization of a key result of \cite{parker1958dynamics}, to find the position of the critical point, and the corresponding mass-loss rate. The equality at the critical point (Eq. \ref{eq: Critical Point Condition}) implies that the specific kinetic, thermal and potential energies of the gas are all of the same magnitude, $k_BT_{\rm{ion}}/\mu\approx\Delta S^{-1}U/\mu$.
Thus, the dominant part of the energy of the gas, assuming it is partially ionized, is the ionization energy. Using Eq. (\ref{eq: Bernoulli Parameter of Transonic Solution}), neglecting all terms except the ionization energy, we determine the ionization fraction at the critical point to first order in $\Delta S^{-1}$,   
\begin{equation}
    \eta_c\simeq\mu\varepsilon/U.
    \label{eq: Ionization Parmeter at The Critical Point}
\end{equation}
For the ionization energy at the critical point to be significant, we implicitly assume that $\varepsilon\gg \Delta S^{-1}U/\mu$, otherwise the critical point would be too close to the non-ionized region and Eq. (\ref{eq: Ionization Parmeter at The Critical Point}) would become inaccurate. 

Using the triple equality at the critical point, Eq. (\ref{eq: Critical Point Condition}), we extract the local velocity and the radius at the critical point,
\begin{align}
    v_c&=\sqrt{\gamma_c\frac{k_BT_c}{\mu}\left(1+\eta_c\right)}\sim\sqrt{\Delta S^{-1}\frac{U}{\mu}},
    \label{eq: Velocity at The Critical Point}
    \\r_c&=\frac{GM}{2c_{s,c}^2}\sim\Delta S\frac{GM}{U/\mu},
    \label{eq: Position of the Critical Point}
\end{align}
where $T_c$ and $\gamma_c$ are determined by substituting $\eta_c$ into Eqs. (\ref{eq: Entropy of Ionized Gas}) and (\ref{eq: The adiabatic index of inozed gas}), respectively. Note that the radius at the critical point is much larger than the recombination radius, i.e. $r_c\gg r_{\rm{rec}}$. 

The density at the critical point is then given by substituting $T_c$ and $\eta_c$ into Eq. (\ref{eq: Density Profile During Adaiabatic Expansion}).
Substituting these results (Eqs. \ref{eq: Density Profile During Adaiabatic Expansion}, \ref{eq: Velocity at The Critical Point} and \ref{eq: Position of the Critical Point}) into Eq. (\ref{eq: mass-loss Rate}), we derive the mass-loss rate of the star due to recombination-driven winds,
\begin{multline}
    \dot{M}=0.5h^{-3}\left(GM\right)^2m_e^{3/2}\left(2\pi\mu\right)^{5/2}\\\times\gamma_c^{-3/2}\left(1+\eta_c\right)^{-3/2}F\left(\eta_C\right)\exp{\left(\frac{5}{2}-\frac{\Delta S}{1+\eta_c}\right)}\\=5.8\cdot10^{7}\left(\frac{M}{M_\odot}\right)^2\gamma_c^{-3/2}\left(1+\eta_c\right)^{-3/2}F\left(\eta_c\right)\\\times\exp{\left(-\frac{\Delta S}{1+\eta_c}\right)}~M_\odot \rm{yr}^{-1}.
    \label{eq: mass-loss Rate of Ionized Gas Wind}
\end{multline}
Note that $\gamma_c$ and $\eta_c$ are functions of the entropy and the Bernoulli parameter themselves, so $\dot{M}$ is solely a function of $S$ and $\varepsilon$.
As long as the ionization fraction is determined accurately, this equation is accurate to all orders in $\Delta S^{-1}$. While the approximation of Eq. (\ref{eq: Ionization Parmeter at The Critical Point}) gives the correct order of magnitude of all physical quantities, Eq. (\ref{eq: Density Profile During Adaiabatic Expansion}) implies that in order to determine the correct numerical coefficient of the density at the critical point (and the mass-loss rate), the ionization fraction has to be calculated at least to first order in $\Delta S^{-1}$. Substituting Eqs. (\ref{eq: Position of the Critical Point}) and (\ref{eq: Velocity at The Critical Point}) into Eq. (\ref{eq: Bernoulli Parameter of Transonic Solution}), and dividing by $U/\mu$, we find an equation for the ionization fraction at the critical point,
\begin{equation}
    \frac{\mu\varepsilon}{U}=\left(\frac{5}{2}-\frac{3}{2}\gamma_c\right)\frac{k_BT_c}{U}\left(1+\eta_c\right)+\eta_c.
    \label{eq: Equation to Determine Ionization Fraction}
\end{equation}
Substituting Eqs. (\ref{eq: Entropy of Ionized Gas}) and (\ref{eq: The adiabatic index of inozed gas}) into the result above, this equation can be solved numerically. Nevertheless, we solve it analytically to second order in $\Delta S^{-1}$. The ionization fraction at the critical point to second order in $\Delta S^{-1}$ is given by
\begin{equation}
    \eta_c=\frac{\mu\varepsilon}{U}-\Delta S^{-1}\left(1+\frac{\mu\varepsilon}{U}\right)^2.
    \label{eq: Ionization Fraction to Second Order}
\end{equation}
Substituting this result into Eq. (\ref{eq: mass-loss Rate of Ionized Gas Wind}), taking the high entropy limit of $\gamma_c\rightarrow1$, we determine the mass-loss rate due to recombination-driven winds to first order in $\Delta S^{-1}$,
\begin{multline}
    \dot{M}\simeq2.1\cdot10^{7}\left(\frac{M}{M_\odot}\right)^2\left(1+\frac{\mu\varepsilon}{U}\right)^{-3/2}F\left(\frac{\mu\varepsilon}{U}\right)\\\times\exp{\left(-\frac{\Delta S}{1+\mu\varepsilon/U}\right)}~M_\odot \rm{yr}^{-1}.
    \label{eq: First Order Mass-Loss Rate}
\end{multline}
Substituting the Sun's mass  into the result above, an entropy per baryon $S=36.0$, and a Bernoulli parameter of $\varepsilon=0.01~U/\mu$, we get a mass-loss rate of $\dot{M}\simeq10^{-3}~M_\odot\rm{yr}^{-1}$. We can use the result of Eq. (\ref{eq: Ionization Fraction to Second Order}) to find all quantities to second order in $\Delta S^{-1}$. We find this calculation sufficient for most quantities (see appendix \ref{app: The Accuracy of The High Entropy Limit} for more details).

Upon reaching the critical point, the local velocity is equal to half of the local escape velocity. In the environment of the critical point, up to radii a few times the radius of the critical point (Eq. \ref{eq: Position of the Critical Point}), the ionization fraction and the temperature of the gas are nearly constant (Eq. \ref{eq: Ionization Parmeter at The Critical Point}), while the velocity increases with $r$, eventually surpassing the escape velocity. Thus, shortly after passing the critical point, the gas becomes unbound, and is able to freely escape to infinity. According to Eq. (\ref{eq: Bernoulli Parameter of Transonic Solution}),
the velocity of the gas approaches $v\rightarrow\sqrt{2\varepsilon}$.

Note that our solution is only valid if the photons released by recombination escape after the gas has reached the escape velocity, which only occurs outside the critical point. Otherwise, the ionization energy is not efficiently converted to the kinetic energy of the gas, but instead it leaks in the form of radiation. Further research is required in order to determine under which conditions the solution is applicable.

\begin{figure*}[t]
    \centering
    \includegraphics[width=0.9\textwidth]{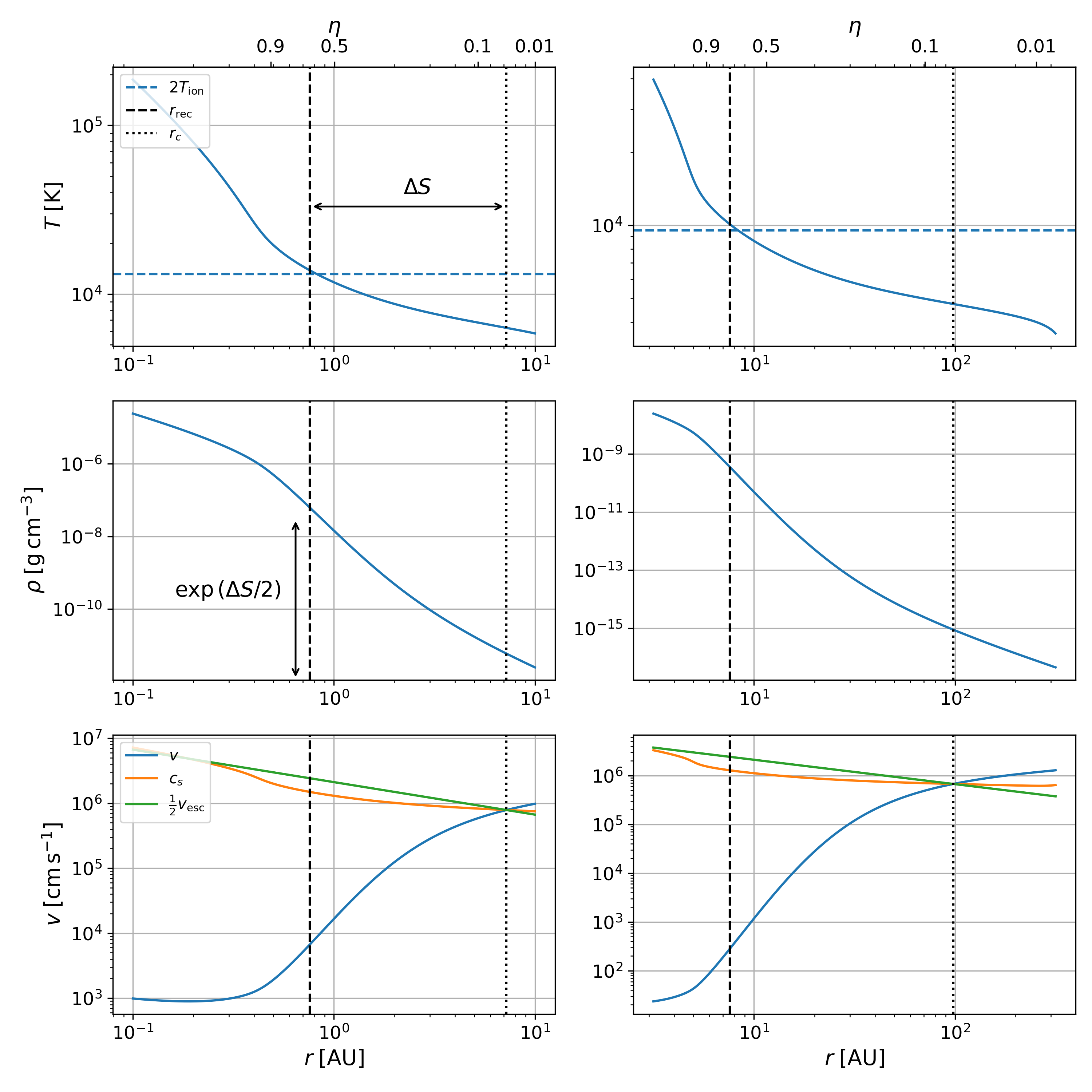}
    \caption{The transonic solution of recombination-driven wind for hydrogen gas derived from the numerical solution to Eq. (\ref{eq: Equation to Determine Ionization Fraction}) (see appendix \ref{app: The Accuracy of The High Entropy Limit} for more details). The left panels correspond to a stellar mass of $1~M_\odot$ and an entropy per baryon $S=36.0$, while the right panels correspond to a stellar mass of $10~M_\odot$ and an entropy per baryon $S=45.0$. Both cases correspond to a Bernoulli parameter of $\varepsilon=0.1~U/\mu$. \textbf{From Top to Bottom}: The temperature (in units of $\rm{K}$, blue horizontal dashed lines show $T=2T_{\rm{ion}}$), density (in units of $\rm{g}/\rm{cm}^3$), and velocities (blue, orange and green lines represent the velocity, sound speed and half of the escape velocity, respectively). All of the physical quantities are presented as a function of the radius (in units of $\rm{AU}$), from deep inside the hydrostatic region to outside the critical point. The recombination radii (at $r_{\rm{rec}}=0.76~\rm{AU}$ and $r_{\rm{rec}}=7.6~\rm{AU}$, given by Eq. \ref{eq: Recombination Radius}) and the critical points (at $r_c=7.2~\rm{AU}$ and $r_c=98.0~\rm{AU}$), are marked with black dashed and dotted lines, respectively. The top axes present the local ionization fraction $\eta$, that decreases with the radius.
    }
    \label{fig: Numerical Transonic Solution}
\end{figure*}

Fig. (\ref{fig: Numerical Transonic Solution}) shows the full transonic solution, from deep inside the hydrostatic region and beyond the critical point, obtained from the numerical solution of Eq. (\ref{eq: Equation to Determine Ionization Fraction}), for two cases (see appendix \ref{app: The Accuracy of The High Entropy Limit} for more details): Entropies per baryon of $S=36.0$ and $S=45.0$, with stellar masses of $1~M_\odot$ and $10~M_\odot$, respectively. Both cases correspond to a Bernoulli parameter of $\varepsilon=0.1~U/\mu$. According to Eq. (\ref{eq: Bernoulli Parameter of Transonic Solution}), at low radii the temperature follows a power law of the radius, $T\propto r^{-1}$, until it drops below $U/k_B$. Then, the slope of the graph becomes steeper, until the temperature reaches $2T_{\rm{ion}}$ at the recombination radius $r_{\rm{rec}}$. It remains of order of $T_{\rm{ion}}$ until the gas fully recombines, a little outside the critical point. It shows that while the temperature remains nearly constant, the density drops by a few orders of magnitude. Additionally, the figure demonstrates the triple equality of Eq. (\ref{eq: Critical Point Condition}), fulfilled at the critical point.

\citet{Quataert_2016} presented stationary transonic solutions for Super-Eddington flows, governed by spherically symmetric Euler equations in the gravitational field of a massive star. Instead of recombination energy they invoke a  general heat source. Their solution also exhibits a transonic point and a Bernoulli parameter that changes according to the heat source. However, while we use an EOS given by Eq. \ref{eq: EOS of Ionized Gas}, \citet{Quataert_2016} uses an EOS of ideal gas, with an adiabatic index of $\gamma=\frac{4}{3}$.

\section{Evolution of Ionized Gas Envelope}
\label{sec: Evolution of Ionized Envelope}
    In \S \ref{sec: The Transonic Solution}, we use three quantities to describe the transonic solution: the Bernoulli parameter $\varepsilon$, the entropy per baryon $S$, and the total stellar mass $M$, all of which are constant along the flow. However, they are not time-independent. Because of the recombination-driven winds, the envelope experiences mass-loss, so the stellar mass $M$ decreases. Moreover, as illustrated by Eq. (\ref{eq: Recombination Radius}), the Bernoulli parameter can be derived directly from the physical quantities of the star, so it is also time-dependent, both because of the decreasing mass and because the deep envelope may have a varying specific entropy. We assume that the gas obeys the EOS of a single-level ionized gas (Eq. \ref{eq: EOS of Ionized Gas}), implying that its adiabatic index is $\gamma=5/3$ in the fully ionized region. Since stable polytropic envelopes (with an adiabatic index of $\gamma>4/3$) expand when experiencing mass-loss \cite{chandrasekhar1939, kippenhahn2012}, as long as their mass is much higher than that of the stellar core, the Bernoulli parameter must increase. In this section, we find a relation between the stellar mass and the Bernoulli parameter. We use this result, together with Eq. (\ref{eq: mass-loss Rate of Ionized Gas Wind}), to calculate the evolution of both quantities as a function of time. Note that the entropy per baryon $S$ remains constant throughout the evolution.

As discussed in \S \ref{sec: Hydrostatic Equilibrium}, in the high entropy limit, the gas is nearly fully ionized for radii lower than $r_{\rm{rec}}$. Outside this radius, we treat the stellar mass $M$ as a constant, and the gravitational potential is approximately given by $\phi\simeq-\frac{GM}{r}$. Below it, the stellar structure is obtained from hydrostatic equilibrium and the polytropic relations, derived from Eqs. (\ref{eq: EOS of Ionized Gas}) \& (\ref{eq: Density Profile During Adaiabatic Expansion}) in the completely ionized gas limit $\eta\rightarrow1$:
\begin{equation}
    P=\frac{h^2\mu^{-5/3}}{\pi m_e}\exp{\left(\frac{\Delta S-5}{3}\right)}\rho^{5/3}.
    \label{eq: Polytropic Relations in The Star}
\end{equation}
Substituting the relation between the recombination radius and the Bernoulli parameter (Eq. \ref{eq: Recombination Radius}) into the well known mass-radius relations of a polytropic envelope (which is much more massive than the stellar core), $RM^{1/3}\simeq K/0.4242G$ \cite{Polytropes_Princeton, chandrasekhar1939, kippenhahn2012}, where $K$ is the coefficient of $\rho^{5/3}$ in Eq. (\ref{eq: Polytropic Relations in The Star}), we obtain the stellar mass as a function of the Bernoulli parameter,
\begin{multline}
    M=0.4242^{-3/4}~h^{3/2}\mu^{-5/4}G^{-3/2}\left(\pi m_e\right)^{-3/4}\\\times\exp{\left(\frac{\Delta S-5}{4}\right)}\left(U/\mu-\varepsilon\right)^{3/4}
    \\\simeq\exp{\left(\frac{\Delta S-24.0}{4}\right)}\left(1-\frac{\mu\varepsilon}{U}\right)^{3/4}~M_\odot.
    \label{eq: Stellar Mass Function}
\end{multline}
For entropies per baryon $S=36.0$ and $S=45.0$ (corresponding to $\Delta S=24.0$ and $\Delta S=33.0$, respectively), when the Bernoulli parameter approaches $0$, i.e. $\varepsilon\ll U/\mu$, Eq. (\ref{eq: Stellar Mass Function}) predicts stellar masses of approximately $1~M_\odot$ and $10~M_\odot$, respectively.

\begin{figure}[hbtp]
    \centering 
    \includegraphics[width=1.0\linewidth]{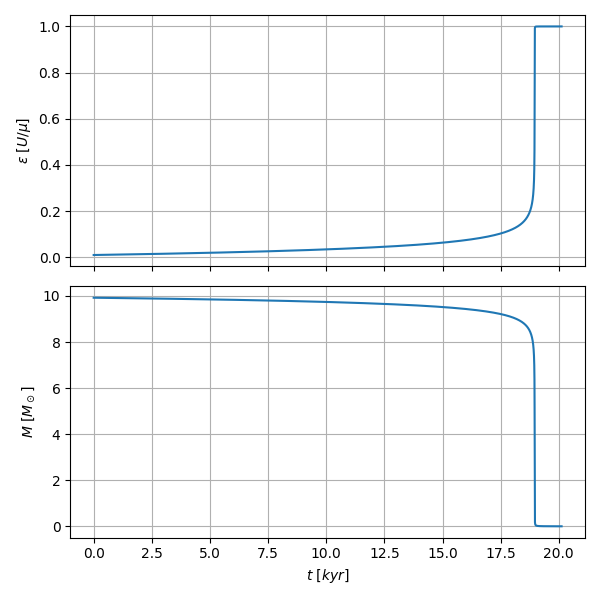}
    \caption{Evolution of a stellar polytropic envelope with entropy per baryon $S=45.0$ and an initial Bernoulli parameter $\varepsilon=0.01~U/\mu$ (corresponding to an initial mass of $\approx10~M_\odot$). \textbf{Top panel}: The Bernoulli parameter, $\varepsilon$, in units of $U/\mu$. \textbf{Bottom panel}: The stellar mass, $M$, in units of $M_\odot$. Both are presented as a function of time, in units of $\rm{kyr}$.}
    \label{fig: Evolution of Stellar Envelope}
\end{figure}

For any stellar mass $M$ and entropy per baryon $S$, we use Eq. (\ref{eq: Stellar Mass Function}) to find the corresponding Bernoulli parameter. Using this result, we integrate over Eq. (\ref{eq: First Order Mass-Loss Rate}) to find the stellar mass $M$ and the Bernoulli parameter as a function of time.
Fig. (\ref{fig: Evolution of Stellar Envelope}) shows this solution for an initial Bernoulli parameter of $\varepsilon=0.01~U/\mu$, given that the entropy per baryon is $S=45.0$, corresponding to an initial mass of $\approx10~M_\odot$.

According to Eq. (\ref{eq: Stellar Mass Function}), as the stellar mass decreases, the Bernoulli parameter increases, asymptotically approaching $\varepsilon\rightarrow U/\mu$. Consequently, the mass-loss rate increases (Eq. \ref{eq: First Order Mass-Loss Rate}), leading to a \emph{Mass Runaway} process. This continues until the envelope's mass becomes comparable to that of the stellar core, causing it to shrink and the Bernoulli parameter to decrease. Eventually, the Bernoulli parameter drops below $0$, terminating the stellar wind. Moreover, Eq. (\ref{eq: First Order Mass-Loss Rate}) implies that the mass-loss rate also decreases when $U/\mu-\varepsilon$ becomes comparable to $\Delta S^{-1}U/\mu$, even if the envelope's mass remains much larger than that of the core. However, since in this regime $\varepsilon\ll U/\mu$, and the stellar mass has already decreased significantly (Eq. \ref{eq: Stellar Mass Function}), we expect the recombination-driven winds to be terminated at an earlier stage.
We evaluate the final envelope mass by calculating the mass of a polytropic envelope surrounding a point core with stellar mass of $M_c$, which is much more massive than the envelope surrounding it, i.e. $M_c\gg M_{\rm{env}}$, given a Bernoulli parameter $\varepsilon$. We substitute $v=0$ and $\eta=1$ into Eq. (\ref{eq: Bernoulli Parameter of Transonic Solution}), with $M\simeq M_c$, to find the temperature as a function of the radius. Following this, we integrate over Eq. (\ref{eq: Density Profile During Adaiabatic Expansion}), in the limit $\eta\rightarrow1$, from the core to the recombination radius, and obtain the envelope mass as a function of the Bernoulli parameter $\varepsilon$:
\begin{multline}
    M_{\rm{env}}\simeq
    \frac{\pi^{7/2}\mu^{5/2}m_e^{3/2}}{2^{1/2}5^{3/2}h^3}\left(GM_c\right)^3\left(\frac{U}{\mu}-\varepsilon\right)^{-3/2}\\\times\exp{\left(\frac{5-\Delta S}{2}\right)}
    \\=\exp{\left(\frac{25.6-\Delta S}{2}\right)}\left(\frac{M_c}{M_\odot}\right)^3\left(1-\frac{\mu\varepsilon}{U}\right)^{-3/2}~M_\odot.
    \label{eq: Final Envelope Mass}
\end{multline}
To find the final envelope mass (at the moment of termination of the recombination-driven winds), we simply substitute $\varepsilon=0$ into the result above. For the case shown in Fig. (\ref{fig: Evolution of Stellar Envelope}), given a stellar core of $M_c=3~M_\odot$, Eq. (\ref{eq: Final Envelope Mass}) predicts a final envelope mass of $\approx0.6~M_\odot$, meaning most of the envelope's mass is ejected in this case. 
We evaluate the time from the onset of the recombination-driven winds until they are terminated by taking the limit $\varepsilon\ll U/\mu$, where Eq. (\ref{eq: First Order Mass-Loss Rate}) becomes 
\begin{equation}
    \dot{M}\approx2.1\cdot10^7\left(\frac{M}{M_\odot}\right)^2\exp{\left(-\Delta S\left(1-\mu\varepsilon/U\right)\right)}.
    \label{eq: Approximated Mass Loss Rate}
\end{equation}
Substituting Eq. (\ref{eq: Stellar Mass Function}) into the above result (again in the limit of $\varepsilon\ll U/\mu$), we evaluate $\dot{\varepsilon}$, and derive the behavior of the Bernoulli parameter at early times:
\begin{equation}
    \varepsilon=-\frac{U/\mu}{\Delta S}\ln{\left(2.8\cdot10^7\left(\frac{M_0}{M_\odot}\right)\Delta Se^{-\Delta S}\left|\frac{\Delta t}{\rm{year}}\right|\right)},
    \label{eq: Bernoulli Parameter at Early Times}
\end{equation}
where $M_0$ is the initial mass. Note that $\Delta t<0$ and its origin are determined using the initial conditions. Thus, the time until the termination of the mass runaway of a star with an initial Bernoulli parameter $\varepsilon_0$ is approximately given by
\begin{equation}
    \tau_{\rm{MR}}\approx\frac{\exp{\left(\Delta S\left(1-\mu\varepsilon_0/U\right)\right)}}{2.8\cdot10^7\Delta S\left(M_0/M_\odot\right)}~{\rm years}.
    \label{eq: Time Until MassRunaway}
\end{equation}
For the case shown in Fig. (\ref{fig: Evolution of Stellar Envelope}), Eq. (\ref{eq: Time Until MassRunaway}) gives $\tau_{MR}=20.0~\rm{kyr}$.

Note that calculating $M/\dot M$ from Eq. (\ref{eq: Approximated Mass Loss Rate}) gives a timescale which is $\Delta S$ times longer than the out estimate in Eq. (\ref{eq: Time Until MassRunaway}). This is because as soon as the star loses a small fraction $\Delta S^{-1}$ of its mass, epsilon increases, and with it the mass loss rate increases.
 
We have treated the evolution of the ionized envelope as a transition from one stationary solution to another since its instantaneous evolution timescale, estimated by $M/\dot{M}$, is much longer than the time required for a fluid element to travel from the recombination radius to the critical point. The latter time scale is $\tau_{flow}\sim M_{flow}/\dot{M}$, where $M_{flow}\ll M_{env}$ is the mass enclosed between the recombination radius and the critical point. Hence, this condition is satisfied, except at the latest stage of the evolution, where the envelope mass is close to its final value given by Eq. (\ref{eq: Final Envelope Mass}).

\section{Recombination-Driven Winds in CEI }
\label{sec: Recombination-Driven Winds in CEI}
In this section, we investigate the link between our recombination-driven winds analytical solution and the phenomenon of CEI. The initialization of CEI has been studied extensively \cite{Paczynski_1976, Ivanova_2013, Nandez_2016}. A binary system of two stars on a Keplerian orbit destabilizes due to Roche-lobe overflow from the donor star to its companion, causing the companion to spiral-in towards the donor. Eventually, the companion becomes fully engulfed by the donor's envelope, and the "common envelope" phase begins. \citet{Ivanova_2015} has explored the role of recombination energy during CEI, showing that the sum of energy deposited by orbital decay and recombination energy of the donor's envelope might lead to a full envelope ejection. Such events may have a nearly steady outflow phase \cite{Nandez_2016, Ivanova_2016}, eventually leading to the development of recombination-driven winds that could be described using our solution.

Initially, the donor, which is assumed to be at its red giant or asymptotic giant phase, possesses a nearly fully ionized and isentropic envelope with a mass $M$ and a finite radius $R$. As the companion spirals inside the common envelope, it heats the envelope, raising its Bernoulli parameter and inflating it. Eventually, the recombination energy per baryon is enough to start unbinding the gas, i.e. the Bernoulli parameter becomes positive. Consequently, recombination-driven winds develop, and the outer part of the common envelope connects smoothly to the hydrostatic part of the wind solution. The mass-loss rate can be estimated by substituting the envelope's new entropy and Bernoulli parameter into Eq. (\ref{eq: First Order Mass-Loss Rate}).

The evolution of the common envelope until the termination of the recombination-driven winds is described in \S\ref{sec: Evolution of Ionized Envelope}. The termination occurs when the envelope's mass becomes comparable to that of the stellar core- or, in the case of CEI, to the total mass of both of the stellar cores. At this point, the mass-loss would cause the envelope to shrink, eventually leading to a negative Bernoulli parameter. Consequently, for systems that have most of their mass is in the envelope, most of the envelope is ejected. 
Note that the energy deposited by the companion may raise the value of the Bernoulli parameter, $\varepsilon$, to significant fractions of $U/\mu$. In this case, the recombination-driven wind will begin near the end of the evolution shown in Fig. (\ref{fig: Evolution of Stellar Envelope}), implying large mass loss rates and short wind durations (Eqs. \ref{eq: First Order Mass-Loss Rate} and \ref{eq: Time Until MassRunaway}). 

\begin{figure}[hbtp]
    \centering
    \includegraphics[width=\linewidth]{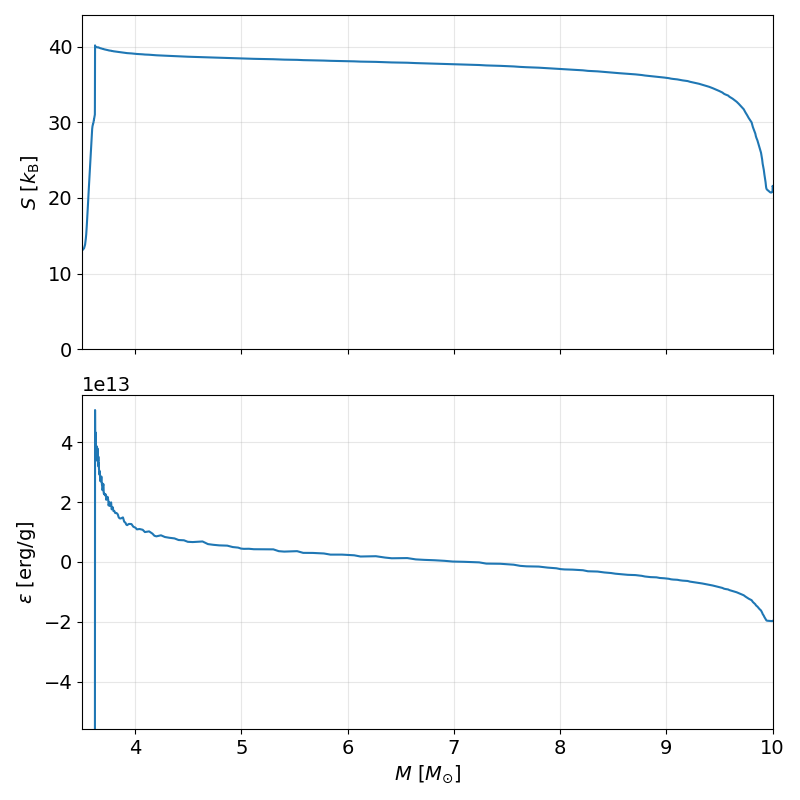}
    \caption{The entropy per baryon $S$, and the Bernoulli parameter $\varepsilon$ (in units of $\rm{erg}\cdot\rm{g}^{-1}$), of a MESA \cite{Jermyn_2023} model of a $10~M_\odot$ red supergiant envelope, as a function of the mass coordinate. Note that the top panel presents a lower entropy than the expected value of $S=45.0$ for a $10~M_\odot$ star given by Eq. (\ref{eq: Stellar Mass Function}), because Eq. (\ref{eq: Stellar Mass Function}) does not account for the helium present in the envelope of a realistic star.}
    \label{fig: Entropy and Bernoulli parmeter of a red giant}
\end{figure}

Although the Bernoulli parameter near the surface of the initial donor envelope must be negative (and roughly equal to $-GM/R$), in some cases it might be positive throughout a significant fraction of the envelope's mass, as shown in the case presented by Fig. (\ref{fig: Entropy and Bernoulli parmeter of a red giant}). In this case, the companion only needs to deposit a small amount of orbital energy in order to trigger the recombination-driven winds and eject the entire common envelope, while its orbit barely shrinks. This process is expected to leave behind post common envelope binaries (PCEB) with wide orbits. Indeed, many such candidates were recently observed by Gaia \cite{Yamaguchi_2024, shariat2026globalviewpostinteractionwhite}.

These results neglect radiative losses in our analysis, which may inhibit the ability of wind ejection \cite{Sabach_2017, chamandy2024negativefeedbackambientenvironment, Grichener_2018, Soker_2018}.

\section{Conclusions and Discussion}
\label{sec: Conclusions and Discussion}
In this work, we analytically derive a stationary solution to the spherically symmetric Euler equations in the gravitational field of a star (appearing in \S \ref{sec: The Transonic Solution}). We use the EOS of a single-level ionized gas (Eq. \ref{eq: EOS of Ionized Gas}) in the limit of high entropy per baryon, $\Delta S\gg1$, and show that in this limit the ionization fraction is approximately linearly dependent on the temperature. In this limit, the ionization fraction increases from roughly zero at the ionization temperature, defined as  $T_{\rm{ion}}=\Delta S^{-1} U/k_B$, to roughly one at $2 T_{\rm ion}$. The flow is accelerated due to the recombination energy release at radii above the recombination radius, where the star's temperature is $2 T_{\rm ion}$.

The solution is transonic and therefore it smoothly connects to a nearly-hydrostatic equilibrium solution of the stellar structure at radii smaller than the recombination radius (Eq. \ref{eq: Recombination Radius}), and to a wind solution with a nearly constant wind velocity at $r\rightarrow\infty$. It describes a partially ionized gas that is accelerated from subsonic to hypersonic velocities at $r\rightarrow\infty$, which implies the existence of a critical point, where the local gas velocity $v_c$ is equal to the local sound speed $c_{s,c}$. In appendix \ref{app: Uniqness of The Transonic Solution}, we generalize the result of Parker \cite{parker1958dynamics} and show that the continuous transonic solution is unique. In \S\ref{sec: The Transonic Solution}, we use this result to calculate the mass-loss rate (Eq. \ref{eq: mass-loss Rate}) of recombination-driven winds. In \S \ref{sec: Evolution of Ionized Envelope}, we develop the evolution of a polytropic stellar envelope made of a single-level ionized gas, and show that the recombination-driven winds lead to mass runaway, which is terminated only when the envelope's mass is comparable to that of the stellar core. We evaluate the final envelope mass, and the time from the initialization to termination of the recombination-driven winds.
In binary star systems, recombination-driven winds may be initiated by energy input to the envelope from a companion star, e.g., during a common-envelope event \cite{Ivanova_2013, Ivanova_2015, Nandez_2015, Ivanova_2018}. For sufficiently extended stars, only a small amount of energy input is needed to raise the envelope's Bernoulli parameter above zero such that a wind can be sustained using recombination energy. This may allow binary systems to eject the envelope of the extended star with only a small amount of orbital decay.

Our work can be generalized for ionized gas which is made of a combination of helium and hydrogen, and may serve as a basis to estimate the mass-loss rate in common envelope interaction, provided that such events would trigger recombination-driven winds. Additionally, once radiation effects are incorporated, it might be used to provide further explanation to the link between CEI and LRNe events \cite{chen2024bridginggapluminousred, Kirilov_2025}. While in this paper, we only deal with the case of a single-level ionized gas wind, this method can be generalized to all cases the gas has an additional source of internal energy, other than the ionization energy \citep[e.g.,][]{Quataert_2016}.
In fact, as we were about to submit this paper, a related paper  appeared \cite{yang2026steadystatestellarwindsdriven}. Compared to our paper, they use a tabulated EOS, including the recombination of helium and accounting for radiative losses. 

\begin{acknowledgments}
This research was partially supported by an NSF/BSF grant, a MOS grant and a GIF grant.
\end{acknowledgments}

\appendix
\section{The Entropy of Hydrogen-Like Ionized Gas}
\label{app: The Entropy of Hydrogen-Like Ionized Gas}
The entropy per baryon (in units of $k_B$) of hydrogen-like gas is the weighted sum of the entropies of the particles the gas is made up of: Atoms in the ground state, ionized atoms, and free electrons. Thus, the entropy of a partially ionized hydrogen-like gas is given by
\begin{multline}
    S = \frac{n_0}{n_0+n_1}\left(\frac{5}{2}+\ln{\left(\frac{2}{n_0\lambda_\mu^3}\right)}\right)\\\frac{n_1}{n_0+n_1}\left(\frac{5}{2}+\ln{\left(\frac{1}{n_1\lambda_\mu^3}\right)}\right)\\+\frac{n_e}{n_0+n_1}\left(\frac{5}{2}+\ln{\left(\frac{2}{n_e\lambda_{e}^3}\right)}\right),
    \label{eq: Entropy per Baryon in The Sackur-Tetrode Formalism}
\end{multline}
where $n_i$ is the number density of atoms in the i'th ionization levels, $n_e$ is the number density of electrons, and $\lambda_{\mu,e}$ are the thermal de-Broglie wavelength of the atoms and electrons, respectively (Eq. \ref{eq: The thermal de-Broglie wavelength} with the corresponding masses). The numerical coefficient inside the logarithm arises from the spin degeneracy of atoms in the ground state and free electrons. Since, by definition, $\eta\equiv n_1/\left(n_0+n_1\right)$, and $n_e=n_1$, we can replace all number densities with $\eta$ and $\rho/\mu\equiv n_0+n_1$. Additionally, we substitute $\lambda_\mu=\left(m_e/\mu\right)^{3/2}\lambda_e$. The result is given by:
\begin{multline}
    S = \left(1-\eta\right)\left(\frac{5}{2}+\ln{\left(\frac{2\mu}{\left(1-\eta\right)\rho\lambda_e^3}\right)}+\frac{3}{2}\ln{\left(\frac{\mu}{m_e}\right)}\right)\\+\eta\left(\frac{5}{2}+\ln{\left(\frac{\mu}{\eta\rho\lambda_e^3}\right)}+\frac{3}{2}\ln{\left(\frac{\mu}{m_e}\right)}\right)\\+\eta\left(\frac{5}{2}+\ln{\left(\frac{2\mu}{\eta\rho\lambda_{e}^3}\right)}\right)\\=\frac{5}{2}\left(1+\eta\right)-2\eta\ln{\eta}-\left(1-\eta\right)\ln{\left(1-\eta\right)}\\+\left(1+\eta\right)\ln{\left(\frac{\mu}{\rho\lambda_e^3}\right)}+S_0,
    \label{eq: Entropy per Baryon with eta and rho}
\end{multline}
with $S_0\equiv1.5\ln{\left(\frac{\mu}{m_e}\right)}+\ln{2}=12.0$. Substituting Eq. (\ref{eq: Density Profile During Adaiabatic Expansion}) into the result above, we arrive at Eq. (\ref{eq: Entropy of Ionized Gas}) for the entropy per baryon of ionized gas.
\section{The Accuracy of The High Entropy Limit}
\label{app: The Accuracy of The High Entropy Limit}
\begin{figure}[hbtp]
    \centering
    \includegraphics[width=\linewidth]{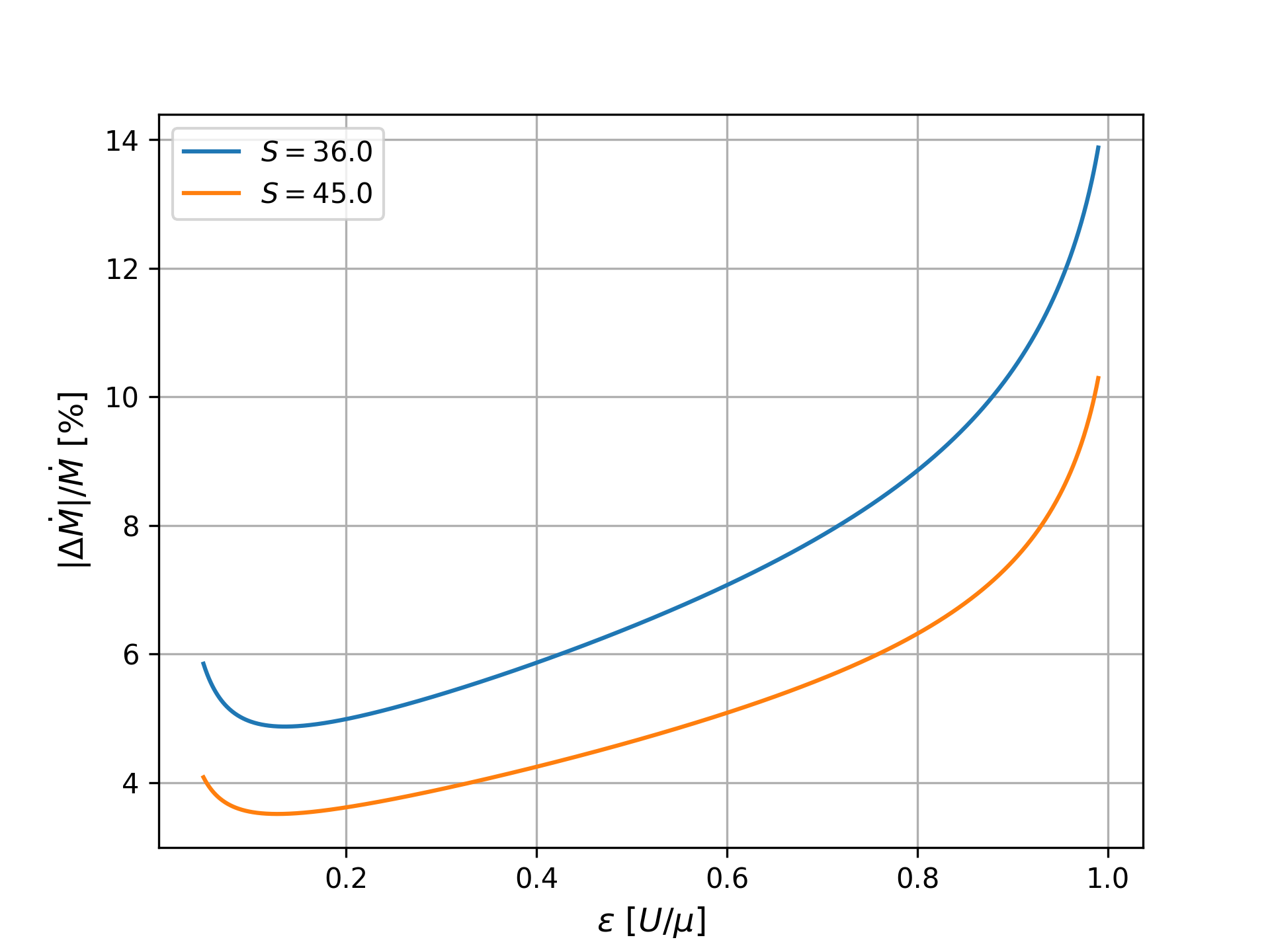}
    \caption{The magnitude of the relative mismatch between the mass-loss rate of Eq. (\ref{eq: mass-loss Rate of Ionized Gas Wind}) to that of the numerical result, as a function of the Bernoulli parameter (in units of $U/\mu$). The blue and orange lines correspond to an entropy per baryon of $S=36.0$ and $S=45.0$, respectively. The magnitude of the mismatch remains below $15\%$ for all Bernoulli parameter values in both cases.}
    \label{fig: Error in Mass-Loss Rate}
\end{figure}

As discussed in \S \ref{sec: The Critical Point}, Eq. (\ref{eq: Equation to Determine Ionization Fraction}) can be solved numerically to find all physical quantities at the critical point to better accuracy than the one obtained analytically by Eq. (\ref{eq: Ionization Fraction to Second Order}) (including the mass-loss rate, given by Eq. \ref{eq: mass-loss Rate of Ionized Gas Wind}). Using the numerically calculated mass-loss rate, we solve Eqs. (\ref{eq: Bernoulli Parameter of Transonic Solution}) and (\ref{eq: mass-loss Rate}) numerically to obtain the full transonic solution everywhere (the results are shown in Fig. \ref{fig: Numerical Transonic Solution}).
We then compare the analytical approximation of \S \ref{sec: The Critical Point} to the numerical solution.
Fig. (\ref{fig: Error in Mass-Loss Rate}) shows the relative mismatch between the mass-loss rate given by Eq. (\ref{eq: First Order Mass-Loss Rate}) and that of the numerical result, in two cases: An entropy per baryon $S=36.0$ and a stellar mass of $1M_\odot$, and an entropy per baryon of $S=45.0$ with a stellar mass of $10~M_\odot$. We perform this calculation for a Bernoulli parameter between $0.05-0.99~U/\mu$. The maximal magnitude of the mismatch remains below $15\%$ for both cases.

Note that the discrepancy between most of the numerically obtained physical quantities at the critical point (such as the ionization fraction, temperature and velocity) and the values obtained analytically using Eq. (\ref{eq: Ionization Fraction to Second Order}), is less than $12\%$. However, the discrepancy between the numerically calculated density at the critical point to that obtained analytically is higher, reaching more than $50\%$ at its peak (for low Bernoulli parameters). Eq. (\ref{eq: Equation to Determine Ionization Fraction}) can be solved to higher orders in $\Delta S^{-1}$ to improve the accuracy of the analytical calculation of the density at the critical point.
\section{Uniqueness of The Transonic Solution}
\label{app: Uniqness of The Transonic Solution}
We are interested in finding the unique stationary solution of the spherically symmetric Euler equations, given in \S\ref{sec: The Transonic Solution}, for a general EOS, that describes the flow of the gas away from a stable massive star of mass $M$. We assume the solution connects smoothly to a hydrostatic solution at $r\rightarrow0$, with a known Bernoulli parameter and specific entropy-- marked as $\varepsilon$ and $S$ respectively-- and to a wind solution at $r\rightarrow\infty$. In the hydrostatic solution regime, the velocity of the gas $v$ is much smaller than the sonic velocity $c_s$, i.e. $v\ll c_s$. However, in the wind solution regime, the velocity of the gas is the dominant one, i.e. $v\gg c_s$. For such a continuous solution to exist, there has to be a critical point, at which $v=c_s$ exactly. We follow a similar procedure as \cite{parker1958dynamics}, and find that demanding regularity at the critical point implies a set of conditions that determine the solution.

We assume the gas flows adiabatically, meaning the specific entropy is conserved along flow lines. Thus, we convert Eq. (\ref{eq: Momentum Conservation Euler Equation}) to
\begin{equation}
    0=v\partial_rv+\frac{c_s^2}{\rho}\partial_r\rho+\frac{GM}{r^2},
    \label{eq: Modified Momentum Conservation}
\end{equation}
where $c_s^2\equiv\left(\frac{\partial P}{\partial\rho}\right)_S$, the sound speed of the gas, is a function of the density $\rho$ and the specific entropy of the gas $S$, derived from the EOS. Using Eqs. (\ref{eq: Mass conservation Euler Equation}) \& (\ref{eq: Modified Momentum Conservation}), we arrive at an ordinary differential equation for the velocity:
\begin{equation}
    0=\left(v-\frac{c_s^2}{v}\right)\partial_rv+\frac{GM}{r^2}-\frac{2c_s^2}{r}.
    \label{eq: ODE for velocity}
\end{equation}
Substituting the equality $v=c_s$ to this result, and demanding the solution to be differentiable at the critical point, implies the existence of a triple equality at the critical point,
\begin{equation}
    v=c_s=\frac{1}{2}v_{esc}(r_c),
    \label{eq: Critical Point Condition}
\end{equation}
where $v_{esc}(r)^2\equiv2GM/r$ is the escape velocity at radius $r$, and $r_c$ is the radius at the critical point. 
As discussed in \S \ref{sec: The Transonic Solution}, the Bernoulli parameter is also conserved along flow lines, and is generally given by
\begin{equation}
    \varepsilon \equiv \frac{v^2}{2}+e+\frac{P}{\rho}-\frac{GM}{r},
    \label{eq: General Bernoulli Parameter}
\end{equation}
where $e$, the specific energy, is a function of $P$ and $\rho$ extracted from the EOS.
Using this result, together with the equality given by Eq. (\ref{eq: Critical Point Condition}), we can find all three variables of the critical point, $r_c$, $v(r_c)$ and $\rho(r_c)$. Substituting these into Eq. (\ref{eq: mass-loss Rate}), we find the mass-loss rate of the star, which together with Eq. (\ref{eq: Bernoulli Parameter of Transonic Solution}), determines the solution everywhere.

As a byproduct of our work, we can use the results of this appendix to restore the properties of the gas at the critical point that obeys a polytropic EOS, with a general adiabatic index $\gamma$ as obtained by \cite{shivamoggi2021parker,shivamoggi2024polytropicgaseffectsparkers}.

\bibliography{references}

\end{document}